\documentstyle[eqsecnum,aps,psfig]{revtex}
\draft

\begin{document}

\title{Relaxation Fluctuations in Quantum Chaos}
\date{\today}
\author{Arul Lakshminarayan \cite{byline} }
\address{ Physical Research Laboratory,
Navarangapura, Ahmedabad,  380009, India.}
\maketitle

\begin{abstract}

Classically chaotic systems relax to coarse grained states of
equilibrium.  This article reports work on quantized
bounded relaxing systems, in particular the quasi-periodic
fluctuations associated with the correlation between two density
operators, whose classical limits as phase space partitions is 
evident. 
These fluctuations can distinguish classically chaotic and regular motions,
thus providing a novel diagnostic devise of quantum chaos. We uncover
several features of the relaxation fluctuations that are shared by
disparate systems thus establishing restricted universality.  
The universal form of these fluctuations for generic quantized chaotic 
systems with time reversal symmetry is worked out within the framework
of Random Matrix Theories and is found to be consistent with the 
numerical simulations performed on varied systems. 

\end{abstract}
\vspace{1cm}
\pacs {To appear in the proceedings of the conference on 
{\sl ``Nonlinear Dynamics and Computational Physics''} held at 
Physical Research Laboratory, Ahmedabad, in Nov. 1997.}

\newpage

\section{Introduction}

\hspace{.5in} Classical dynamical systems have been classified into a
hierarchy
 of deterministic randomness. We have well studied examples
from integrable systems to Bernoulli systems, from the regular to
those that are in a coarse grained manner identical to stochastic
processes.  An important notion in this context is that of mixing, as
exemplified in the now classic coke-rum mixture of Arnold and
Avez \cite{ArnAvez}. Apart from the fundamental role it 
plays in deterministic
randomness, alias chaos, it is the backbone of the foundations of
classical statistical mechanics. While ergodicity is compatible with
equilibrium, it is mixing that would ensure the drive to this state.
The role of deterministic chaos in statistical mechanics is under active
research, for instance it is not clear that thermal equilibrium is 
the result of only deterministic chaos or whether large system 
sizes are a neccessity. 

Low dimensional completely chaotic classical systems are however
characterized by ``equilibrium states'' as the underlying invariant
densities and mixing drives sufficiently well behaved initial states
to this equilibrium. This affords the evaluation of averages much the
same way as in equilibrium statistical mechanics. On the introdcution
of quantum mechanics however, the picture is not quite so clear even
for low dimensional systems.  The phenomenon of quantum suppression of
classical chaos in diffusive systems is now well studied, and several
localization mechanisms have been put forward which inhibit quantum
mixing \cite{Izrailev}, yet the issues in the situation of ``hard
chaos'' for bounded chaotic systems relaxing to an equilibrium have
not been sufficiently addressed.

This paper studies quantum objects that are obviously connected to the twin
issues of ergodicity and mixing in the classical limit. In the process 
we use models having a discrete spectra whose classical limit is known to
be chaotic. The most convenient for our purposes is the quantization 
of two dimensional area preserving maps on the torus- a much studied subject.
The Hilbert space is then finite dimensional and we have fully all the 
contradictions between quantum and classical chaos, while retaining the 
attractive feature of being able to do the quantum mechanics up to machine
precision and size. 
 
Let the relevant classical phase space be $\Omega$. This could be for
instance the energy shell on which the Hamiltonian flow is restricted.
We may formulate mixing either in terms of phase space functions or
densities. Let $A$ and $B$ be two subsets of the phase space such that
they do not intersect. Let $\chi_{A}(x)$ and $\chi_{B}(x)$ be the two
characteristic functions of these subregions. Let the flow integrated
for a time $t$ be denoted by $f^{t}$, this could be a continuous flow
or a discrete map, we formulate this for the case of maps in which
case $t$ is an integer. The invariant measure of the flow is denoted by
$\mu$, in the present context of area preserving maps this is simply the 
physical area of the phase space region. 

The central quantity of interest is the correlation between the
functions $\chi_{A}$ and $\chi_{B}$. In more graphic language, as the
subregion $A$ evolves with the motion the correlation is the
fractional area of its intersection with the region $B$. If in the
long time limit this factorises into the fraction of the areas of $A$
and $B$ we have mixing. In other words the fraction of $A$ 
systems in $B$
is the fractional area of $B$ in the long time limit. 
\begin{equation}
\mu(f^{t}(A) \cap B)/\mu(A) \longrightarrow  \mu(B)/\mu(\Omega)
\end{equation}
Mixing systems are a step above ergodic ones
in the hierarchy of classical dynamical systems. Ergodicity is the equality
of the time average and phase space average of almost all points in the 
phase space. This can also be formulated as decorrelation on the average.
Thus a system is ergodic if
\begin{equation}
 \lim_{T \rightarrow \infty} \frac{1}{T} \sum_{t=1}^{T} \mu(f^{t}(A)
\cap B)/\mu(A) \, =\, \mu(B)/\mu(\Omega).
\end{equation} 

Classical mixing systems are characterized by a series of complex
numbers or resonances that dictate the rate of decay of
correlations. Thus for purely hyperbolic or Axiom A systems we can
write
\begin{equation}
\mu(f^{t}(A) \cap B)/\mu(A)-  \mu(B)/\mu(\Omega)=\sum_{i} C_{i} \exp(\lambda_{i}t). 
\end{equation}
The $\lambda_{i}$ ( Ruelle resonances ) are in
general complex numbers with negative real parts independent of the
particular partitions $A$ and $B$. Explicit 
calculation of these
resonances is a challenging part of dynamical systems theory; there
are special models which are isomorphic to finite Markov processes for
which these can be analytically found; for example the multi-baker
maps \cite{ElsKap,GaspJsp92}.

What we wish to study below are quantities in quantum mechanics that
are naturally related to the classical correlation functions defined
above. These are of course correlations between operators and when the 
dynamics is classically chaotic we expect the quantum correlations to 
also ``decay'' in some sense.  Since we will consider systems with 
discrete spectra this does not strictly happen and eventually there
are oscillations. We will characterize these oscillations and show that
when the dynamics is classically chaotic they are well predicted
by Random Matrix Theory (RMT). 

\section{The Quantum Correlation Function}

The relevant dynamical space on quantization is a Hilbert space
$H_{N}$. For regions of phase space we consider subspaces of the
Hilbert space. Since we have in mind classical maps on the two torus,
let the Hilbert space be of finite dimensionality $N$.  Let there be
two distinct density operators on $H_{N}$ (may or may not be projection
operators), $P_{A}$ and $P_{B}$.  We will study the cases when these
operators on $H_{N}$ have obvious classical limits as functions on
the phase space.  We will restrict ourselves initially, and largely,
to the case when the operators are projection operators such that
$P_{A}+P_{B}=I_{N}$, where $I_{N}$ is the $N$ dimensional identity
matrix. Classically this corresponds to choosing the partitions $A$
and $B$ such that $A \cup B=\Omega$.

We sidestep the question of assigning an operator to a general
subregion of phase space, and whether this is possible at all (say
through the Wigner-Weyl transform of the characteristic function), by
restricting this study to particularly simple subspaces (or density operators)
both of $\Omega$ and on $H_{N}$ and going from the quantum to the
classical instead of vice versa.  Choosing an ``arbitrary '' subspace
( or operator) of $H_{N}$ obviously does not correspond to a proper
subregion of the classical phase space and we will see that such
subspaces do not provide interesting relaxation behaviors in that
they do not distinguish between classically chaotic and regular
motions. 
  
If we use pure states, say  $|n \rangle $, we can  construct the 
density operators as $\sum |n \rangle \langle n  | $.
Let the dimensionality of the subspace $A$ be $n_{A}=f_{A}N$; then
$n_{B}= N-n_{A}=f_{B}N$. The fractions $f_{A}$ and $f_{B}$ determine
the relative sizes of the partitions. 
We will mostly use
projection operators that are diagonal in the position basis. In this
case the classical region corresponding to 
\begin{equation} P_{A}= \sum_{n=0}^{f_{A}N-1} |n \rangle \langle n|
\label{proj-op}
\end{equation} 
is the rectangle
$[0,f_{A}) \times [0,1)$ and the region corresponding to $P_{B}$ is
the rectangle $[f_{A},1) \times [0,1)$, as the phase space is taken to
be the unit torus $(q,p)=[0,1) \times [0,1)$.

The dynamics is an unitary operator $U$ acting on the
states of $H_{N}$. The central quantity of interest is then
\begin{equation}
C_{AB}(t)=\mbox{Tr}(U^{t}P_{A}U^{-t}P_{B})/N,
\end{equation}
 which gives us the overlap of the subspace $A$ propagated for time
$t$ with the subspace $B$. We may note that this is an important and
natural physical quantity to study and has appeared before in several
contexts, for example ref. \cite{UBT}. It can also be viewed as an 
analogue of Landauer conductance
for bound systems and as such should be of considerable importance in
quantum transport.  If in fact there is factorization in time then
$C_{AB}(\infty)$ must be compared with
$\mbox{Tr}(P_{A})\mbox{Tr}(P_{B})/N^{2} = f_{A}f_{B}$. 

Since $U$ has a discrete and finite spectrum, the correlation can only
be a finite sum of purely oscillatory terms and can therefore display
decay only over short time scales (of the order of Heisenberg
time). The decorrelation is therefore not expected. As an extreme
example if we consider the case when the projection operators
are constructed out of the basis functions of $U$ the correlation is
zero for all time. This is an extreme case and for instance it is not
clear what the classical limit is of the Wigner-Weyl transform of 
projection operators
constructed out of such a basis.  We will consider ``generic''
subspaces and projection operators and any claim of universality made
below has to be viewed with this caveat. We may note that this plagues
Random matrix theory descriptions of eigenfunctions as well, as it
involves basis dependent quantities. Here the basis dependence enters
as the basis in which the projection operators are diagonal.

It was suggested earlier using quantized multi-baker maps that the
quantum correlation function approached the classical correlation
function in the classical limit corresponding here to $N \rightarrow
\infty$ \cite{LBJsp}.  Certain remarkable features of quantum
relaxation were noted there including relaxation localization and
effects of symmetries on transport. Here we wish to study the
fluctuation properties more closely and uncover possible universal
features. Therefore we study rather standard models like the
Taylor-Chirikov (standard) map \cite{Izrailev}, the kicked Harper
system \cite{Leb} and the baker map \cite{BVbak} qalthough most of
these models do not have explicit expressions for the classical Ruelle
resonances and are probably not Axiom A systems (except the baker
map). However it is very reasonable to assume that in certain well
known parameter regimes they can be for all practical purposes mixing
systems.

Normalizing the correlation so that the classical limit, if it exists,
is for large times unity we study the quantity
$c(t)=C_{AB}(t)/f_{A}f_{B}$, where from now we will acknowledge
implicitly the dependence of the correlation on the particular choice
of partitions. Fig. 1   shows an example of the generic behaviour of
$c(t)$, this particular data being for the standard map at two
different values of the inverse Planck constant $N$ for the same
classical value of the chaos parameter. The initial relaxing behaviour
is not clear from the figures as the time scales are much larger than
the inverse of the principal Ruelle resonance. The projection
operators used are diagonal in the discrete position basis, $P_{A}$ is
given by Eq.( \ref{proj-op})
and $P_{B}=I_{N}-P_{A}$. The position eigenbasis is denoted as $|n \rangle $
and in this figure $f_{A}=f_{B}=1/2$. 

We have 
used dimensionless scaled position and momentum co-ordinates and 
time is measured as multiples of the period of the kick, taken as unity. 
Modulo one conditions restrict the phase space to the unit torus so that 
the dimension of the Hilbert space $N$ is related to the Planck constant as 
$N=1/h$ and the classical limit corresponds to $N \rightarrow \infty$
while the parameters of the map such as the  kick strengths 
are scaled and dimensionless.

The first observation is that the quantum correlation is in fact quite
close to the classical value of unity, second is that the average of
the oscillations in relation to unity will give us some information
about whether quantum mechanics is inhibiting transport or otherwise,
thirdly this average must in the classical limit tend to unity, fourth
is the observation that the fluctuations are getting smaller in the
classical limit and must tend to zero. Some of these observations have
been made and substantiated earlier \cite{LBJsp,LakPRE}, here we will
elaborate and present more results.

Previous work on the use of time developing states to study the
quantum manifestations of chaos have basically concentrated on the
survival probability of a pure state \cite{SurvProb}. This corresponds
to the choice $P_{A}=P_{B}=|\psi \rangle \langle \psi |$ with an
arbitrary normalized initial state $|\psi \rangle$ and thus
$f_{A}=f_{B}=1/N$. Such survival probabilities averaged over random
initial states show a distinct difference between chaotic and regular
systems. In contrast what we study in this paper are density (or
projection) operators and not pure states. It becomes quite important
that $f_{A}N$ and $f_{B}N$ are large and are of the order of the
Hilbert space dimensionality $N$ itself. Besides any ``arbitrary''
state can be chosen for the calculation of the survival probability
while we will restrict ourselves to those projection operators that
can be interpreted as phase space regions in the classical limit.
This is to ensure that the classical transition to chaos is fully
reflected in the quantum relaxation, as will be illustrated below.

\section{The Fluctuation Properties}

The quantum ``equilibrium'' as opposed to the classical is an highly
oscillatory state, with the fluctuations coming from the discrete
nature of the spectrum. We will denote the time average of the
correlation function by $<c>$ and its variance by $\sigma^{2}$. These
quantities can be written in terms of the eigenfunctions of the
evolution operator $U$, which satisfy the following equation.
\begin{equation} 
 	U|\psi_{m} \rangle = \mbox{e}^{i E_{m}} |\psi_{m} \rangle \; \; 
m=0,\ldots N-1
\end{equation}
The real numbers $E_{m}$ are the eigenangles of the quantum map, and we 
are assuming that there are no exact degenerecies as is the generic case.  
The correlation is then
\begin{equation}
c(t)=\frac{1}{N f_{A} (1-f_{A})} \sum_{m_{1},m_{2}} \sum_{n_{1}=0}^{
Nf_{A}-1} \sum_{n_{2}=Nf_{A}}^{N-1} \mbox{e}^{i t (E_{m_{2}}-E_{m_{1}})} \\
\langle n_{2}|\psi_{m_{1}} \rangle \langle \psi_{m_{1}}|n_{1} \rangle 
\langle n_{1} |\psi_{m_{2}} \rangle \langle \psi_{m_{2}}|n_{2} \rangle 
\end{equation}
Thus we can express the average quite simply as 
\begin{equation}
<c>=\frac{1}{N f_{A}(1-f_{A})} \sum_{m=0}^{N-1} \sum_{n=0}^{f_{A}N-1}
\sum_{n^{\prime}= f_{A}N}^{N-1} |\langle n|\psi_{m} \rangle |^{2}
|\langle n^{\prime}|\psi_{m} \rangle |^{2}.
\end{equation}

Thus the average is a measure of the distribution of the eigenvectors
in the subspaces $A$ and $B$. We suggest that $<c> \, < \, 1$
generically, indicating a certain reluctance to participate equally in
both the partitions in proportion to their sizes.  
A quantity of natural interest is the variance of the fluctuations which
exist irrespective of the symmetries of the system. This can also be 
expressed in terms of the eigenfunctions of the system. In the particular 
case when $P_{A}+P_{B}=I_{N}$, writing for $f_{A}$ simply $f$,
we get after some simplifications 
\begin{equation}
\sigma^{2}= \frac{2}{N^{2} f^{2}(1-f)^{2}} 
\sum_{m_{2}>m_{1}} \left | \sum_{n=0}^{fN-1} \langle n|
\psi_{m_{2}} \rangle \langle \psi_{m_{1}}|n \rangle \right|^{4}.
\end{equation}
Thus the variance measures a correlation between distinct 
pairs of eigenfunctions. There is a coherent partial sum over $Nf$ 
states that makes the variance non-trivial.

\subsection{The Models}

Apart from time reversal symmetry there may be phase space symmetries
that are either preserved or not in the quantum mechanical models.
The results in \cite{LakPRE} are for the case when there is an 
additional phase space symmtery present which is reflected quantally
as a parity symmtery of the eigenfunctions. This being a special (but 
important) case we continue our study here for the more generic
case when the eigenfunctions possess no symmetries. Then the only important
symmtery is that of time reversal.

The Taylor-Chrikov, or the standard map, or the kicked rotor, used in
this paper is described below \cite{Izrailev,ShiChang}.  Let the
classical map be $(q_{t+1}=q_{t}+p_{t+1}, \; \;
p_{t+1}=p_{t}-V^{\prime}(q_{t+1}))$, both $q $ and $p$ taken mod 1.
$V^{\prime}$ is the derivative of the kicking potential which is
assumed to have unit periodicity. The toral states are assumed to
satisfy certain boundary conditions specified by a point on the dual
torus.  Let $|q_{n} \rangle $ and $|p_{m} \rangle $ be the position
and momentum states then $|p_{m+N} \rangle =e^{-2 \pi i a } |p_{m}
\rangle $ and $|q_{n+N} \rangle =e^{2 \pi i b } |q_{n} \rangle $,
where $(a,b)$ are real numbers between 0 and 1; $N$ is the
dimensionality of the Hilbert space. If $b=0$, upon canonical
quantization we get the finite unitary operator
\begin{equation} 
U_{n, n^{\prime}}= \frac{e^{i \pi/4}}{\sqrt{N}} \exp (-2 \pi i N
V(\frac{n+a}{N})) \exp(i \frac{\pi}{N}(n-n^{\prime})^{2}).
\end{equation}
We have used for the standard map $V(q)=K \cos(2 \pi
q)/(2 \pi)$, and $a=1/2$ makes the quantum map possess an exact
parity symmetry about $q=1/2$. Thus we break the symmetry by simply
choosing different boundary conditions for the wavefunctions. If 
$b \ne 0$ then time reversal symmtery is also broken. Therefore
there are two classes of models with no special wavefunction symmetries.  
We call the case $(a=1/2,b=0)$ the symmteric standard map.
The values of $N$ are restricted to the 
even integers.

The kicked Harper map \cite{Leb} on the torus is similar to the above system
except for the momentum dependence. If the Hamiltonian is
\[ H=V_{1}(p)\, +\, 
V_{2}(q) \sum_{n=-\infty}^{\infty} \delta(t-n), \] the quantum map 
is 
\begin{equation} 
U_{n, n^{\prime}}= \frac{1}{N} \exp(-2 \pi i N
V_{2}(\frac{n+a}{N})) \sum_{m=0}^{N-1} \exp(-2 \pi i N V_{1}(\frac{m+b}{N}))
\exp(\frac{2 \pi i}{N} (m+b)(n-n^{\prime}))
\end{equation} 
For the map used below we have taken $V_{1}(p)=-g_{1}\cos(2 \pi p)/(2
\pi)$ and $V_{2}(q)=-g_{2}\cos(2 \pi q)/(2 \pi)$. If $(a=b=1/2)$, this
once again ensures special symmetry properties of the quantum map, a
case we call below the symmetric Harper map. 
We can retain time reversal and break parity by choosing $b=1/2$ and 
$a \ne 1/2$, and or break both symmetries. 
The
classical map is $( q_{t+1}=q_{t}+g_{1} \sin(2 \pi p_{t+1}), \; \,
p_{t+1}= p_{t}-g_{2} \sin( 2 \pi q_{t}))$, again the mod 1 rule is
assumed.  If $g_{1,2}$ are equal, as assumed in this paper, the
transition to chaos occurs around $g_{1,2}=0.63$. In both the kicked
Harper model and the rotor the time between kicks has been taken as
unity, as anyway there are two parameters in the quantum problem, the
scaled Planck constant $N$ and the kick strength ($K$ or $g$).

The bakers map \cite{BVbak,Sarabak} comes in several varieties. 
It has been shown that the
usual bakers map $((q_{t+1}=2 q_{t}, \; p_{t+1}=p_{t}/2)\; \;  \mbox{if}
\; \;  0<q_{t}<1/2 \; \; \mbox{and} \; \; (q_{t+1}=2 q_{t}-1, \; p_{t+1}=(p_{t}+1)/2)
\; \; \mbox{if} \; \; 1/2<q_{t}<1 )$ is isomorphic to the $(1/2,1/2)$
Bernoulli process. The quantum bakers map on the torus is then the
unitary operator
\begin{equation}
U=G_{N}^{-1} \,  \left( \begin{array}{cc}
                        G_{N/2}&0\\
                        0&G_{N/2}
                        \end{array}
                        \right),
\end{equation}
where $G_{N}$ is the finite Fourier transform matrix with elements
\[
(G_{N})_{m,n}=\frac{1}{\sqrt{N}} \exp(-2 \pi i (m+1/2)(n+1/2)/N).
\]
We have assumed anti-periodic boundary conditions $(a=b=1/2)$ and this
is known to preserve classical symmetries of the usual bakers map.

The generalized bakers map used in this paper is a dynamical system
implementing the $(2/3,1/3)$ Bernoulli process and is the classical
map $((q_{t+1}=3 q_{t}/2, \; p_{t+1}=2p_{t}/3) \; \; \mbox{if} \; \;
0<q_{t}<2/3 \; \; \mbox{and} \; \; (q_{t+1}=3 q_{t}-2, \;
p_{t+1}=(p_{t}+2)/3) \; \; \mbox{if} \; \; 2/3<q_{t}<1 )$. The quantum
map is the unitary operator
\begin{equation}
U=G_{N}^{-1} \,  \left( \begin{array}{cc}
                        G_{2N/3}&0\\
                        0&G_{N/3}
                        \end{array}
                        \right).
\end{equation}

\section{Numerical Results}

Time dependent
quantum chaotic systems such as the kicked rotor on the cylinder are
known to suppress classical chaos and lower diffusion \cite{Izrailev}. The
corresponding suppression in the case of bounded systems could be the
average relaxation such as measured by $<c>$.  
We note that the sum over $n$ and $n^{\prime}$
expressing the average can be factored into two single sums, and we
see that if for instance we had a parity symmetry forcing the
wavefunction to be essentially identical in the subspaces $A$ and $B$
we would have $<c>=1$, which is indeed the case for the fluctuations
shown in Fig. 1.  The average $<c>$ could also be greater than unity
in the presence of symmetries, or if the partition $B$ is identical to
$A$.

Fig. 2a  shows the average for the quantized baker map implementing the
Bernoulli scheme $(2/3,1/3)$ as a function of the inverse Planck
constant $N$. The partitions are such that $f_{A}=f_{B}=1/2$. Shown is
the deviation of the average from unity and an approximate power law
is observed for large $N$. Thus we can write
\begin{equation}
<c> \, \sim \, 1-\alpha N^{-\gamma},
\end{equation}  
and in this case $\gamma \approx 3/4$. The relaxation localization is
implied by $\alpha$ being positive.  
Fig. 2b shows the same quantity for the symmetric standard map and we find 
for the partition $f_{A}=1/4, f_{B}=3/4$ that $\gamma \approx 1$,
with a different value of $\alpha$. We note that in both the cases there
is a small oscillation about the fitted straight line. Further results,
especially on the variation of the average with the parameters of chaos,
are in ref. \cite{LakPRE}.

Fig. 3 shows the standard deviation $\sigma$ scaled to $\sigma
N$ as a function of the kick strength in the case of the
symmetric standard map. The transition to classical chaos at $K \approx 1$ is
visible in the variance of the fluctuations as a point at which it
attains an approximately constant value. The relaxation fluctuations
are therefore significantly different depending on the dynamical
nature of the classical limit, being in general larger for regular
systems than chaotic ones, and thus provide a novel diagnostic devise
for the study of quantized chaotic systems. Further results may be found 
in ref. \cite{LakPRE}. Here we will turn to a more closer appraisal of the
ingredients of the variance itself. 

In a slight modification of the equation for the variance, 
define
\begin{equation}
S = 2 \sum_{m_{2}>m_{1}} \left | \sum_{n=n_{0}}^{n_{0}+l-1} \langle n|
\psi_{m_{2}} \rangle \langle \psi_{m_{1}}|n \rangle \right|^{4}.
\end{equation}
The quantity $S$ is simply related to the variance of the
fluctuations. The relevant partitions have been generalized slightly
so that $P_{A}=\sum_{n=n_{0}}^{n_{0}+l-1} |n \rangle \langle n| $ and
the classical partition is the vertical rectangle of length $l/N$
whose origin is at $n_{0}/N$. The $B$ partition is the complementary
space.  $S$ is potentially a function of $l$, $N$ and $n_{0}$. For
generic systems the origin $n_{0}$ should not play a role and we will
expect this dependence to drop out. In fact the case of the symmetric
standard map when the partition is constructed from the position basis
is a counter-example to this statement. But what we will go on to
establish is that for generic quantum chaotic systems and generic
partitions, $S$ is asymptotically (large $N$) only a function of
$f=l/N$ and that the functional dependence is determined by time
reversal symmetry.

In Fig. 4a is plotted the behaviour of $S$ for the symmetric standard map and 
the dependence on $n_{0}$ is clearly visisble as the non-analyticity as a function 
of the fraction $f$. On the other hand Fig. 4b illustrates a smooth functional
dependence independent of $n_{0}$ when the parity symmetry is broken by choosing 
$a=0.35$ instead of $0.5$. The universality of this curve has been checked against
other models such as the kicked Harper map. In the next section a form of this 
smooth function is derived within the framework of RMT.   
Fig. 5 consolidates three cases of the kicked Harper map. The function $S$ with 
both parity (R- symmetry) and time-reversal, with just time-reversal and the 
case when there is neither parity nor time-reversal symmetry. The fluctuations 
clearly distinguish these situations and illustrates the intuitive expectation 
that the fewer the symmetries the smaller would be the fluctuations about the
``quantum equilibrium''.  
 
\section{Fluctuations According to RMT}
 
Random Matrix Theory \cite{Pandey} has been quite successful in the description of many aspects of 
complex quantum systems. It is natural to expect that the fluctuations be also described
adequately within this theory. Already  work on the survival probability has been 
dealt using RMT with success. The fluctuations studied above are characterized by certain
higher order correlations between eigenfunctions and the evaluation of these within
the RMT framework for the case of time reversal symmetric generic systems is
dealt with here.   

The appropriate ensemble of matrices is the circular orthogonal ensemble, but we 
will equivalently use the Gaussian orthogonal ensemble (GOE). Thus in the spirit of
RMT (and statistical mechanics) we will equate any quantity with its ensemble average.
Thus when Tr$(P_{A})=l=fN$ we get
\begin{equation}
<c> = << \, \frac{1}{N f(1-f)} \sum_{m=0}^{N-1} \sum_{n=0}^{l-1}
\sum_{n^{\prime}= l}^{N-1} |\langle n|\psi_{m} \rangle |^{2}
|\langle n^{\prime}|\psi_{m} \rangle |^{2} \, >>.
\end{equation}
The double bracket represent average over the ensemble. GOE for eigenfunctions is particularly
simple with the assumption that each vector covers an $N$ dimensional
sphere uniformly. More details that are crucial to the following can be found for 
instance in \cite{Pandey}. 

Thus under the assumption that there are no favoured eigenfunctions we get 
\begin{equation}
<c> = \frac{l(N-l)}{f(1-f)} << \,|\langle n|\psi_{m} \rangle |^{2}
|\langle n^{\prime}|\psi_{m} \rangle |^{2} \, >>   
\end{equation}
We change to a  different notation purely for the sake of
convenience. We denote the the $n-th$ component of the vector $\lambda$ by
$x_{n \lambda}$. Then using the result \cite{Pandey}
\begin{equation}
<<\, x_{i\lambda}^2 x_{j \lambda}^2 \, >> = \frac{1}{N(N+2)}
\end{equation}
we get
\begin{equation}
<c>\,=\,  \frac{N}{N+2} \sim 1-2/N.
\end{equation}
The final approximation is for large values of $N$ (the classical limit). 
We see that this is consistent with our simulations for the average value of 
the fluctuations. It predicts that for generic systems the average is
less than unity and that the leading correction is proportional to $\hbar$. 
This agrees with our result for the standard map but the bakers map
has a different behaviour as the leading correction was found to be of 
order $\hbar^{3/4}$. This is consistent with what has been observed by others 
so far: that the quantum bakers map is somewhat ``less random'' than other models
of quantum chaos \cite{SteveHeller}. 

The variance is calculated in a similar manner, but leads to higher order
correlations that appear to be a pure ``two-vector'' result in comparison to
that used for the average. Any two eigenvectors of the ensemble are correlated due to the
requirement of orthonormality and the full derivation of the reduced densities is 
not known to the author at present. Nevertheless much can be derived about the 
variance. Concentrating on $S$ (which is trivially related to the variance) we get
\begin{equation}
S= << \, 2 \sum_{m_{2}>m_{1}} \left | \sum_{n=n_{0}}^{n_{0}+l-1} \langle n|
\psi_{m_{2}} \rangle \langle \psi_{m_{1}}|n \rangle \right|^{4} \, >>
\end{equation}
or
\begin{equation}
S= N(N-1)<< \, \left | \sum_{n=1}^{l} x_{n \lambda} x_{n \mu} \right |^4 \, >>
\end{equation}
We have made use of the fact that GOE eigenvectors are real and we also require in the above
that $\lambda \ne \mu$. We have also dropped the $n_{0}$ term (or set it equal to unity)
as the statistical theory does not distinguish between eigenvectors. It is clear therefore
that $S$ is a measure that may be called ``incomplete orthonormality'', while the average is
related to ``incomplete orthogonality''.  

The sum over $l$ terms is equal (due to orthonormality) to a sum over $N-l$ terms and 
therefore statistically we derive that 
\begin{equation}
S(l)=S(N-l)
\end{equation}
where we have explicitly written the dependence of $S$ on $l$. 
We must note that this is not an exact condition of symmetry (from detailed dynamics) 
as is apparent from
the numerical results, but is exact within the framework of RMT. The deviations from
this symmetry could therefore provide an interesting measure of deviation from statistical
behaviour. 

Further expansion of the fourth power in the equation of $S$ and book-keeping leads to
the following fourth order polynomial for $S$. 
\begin{equation}
\frac{S(l)}{N(N-1)}= a_{1} l + a_{2} l(l-1) + 6 a_{3} l(l-1)(l-2) + a_{4} l(l-1)(l-2)(l-3)
\end{equation}
and	
\begin{equation}
\begin{array}{lcl}
a_{1}&=& <<\, x_{i \lambda}^4 x_{j \lambda}^4 \, >>, \\
a_{2}&=&3 <<\, x_{i \lambda}^2 x_{j \lambda}^2 x_{i \mu}^2 x_{j \mu}^2 \, >> +
4 <<\, x_{i \lambda}^3 x_{j \lambda}^3 x_{i \mu} x_{j \mu} \, >>, \\
a_{3}&=&<<\, x_{i \lambda}^2 x_{j \lambda}^2 x_{i \mu} x_{j \mu} x_{i \nu} x_{j \nu}\, >>, \\
a_{4}&=&<<\, x_{i \lambda} x_{j \lambda} x_{i \mu} x_{j \mu} x_{i \nu} x_{j \nu}\,x_{i \alpha} 
x_{j \alpha}\, >>. 
\end{array}
\end{equation}

These are truly two-vector averages ( $i$ and $j$ indices can be also thought
of as denoting vectors rather than components). None of the indices must be the same in each of the above
expressions.
We see quite easily that 
\begin{equation}
a_{4}=3a_{3}/(3-N).
\end{equation}
This follows on summing over the index $\lambda$ in the expression for $a_{4}$ and using orthonormality.
Further we have that 
\begin{equation}
(N-1) << \, x_{i \lambda}^3 x_{j \lambda}^3 x_{i \mu} x_{j \mu} \, >> = - << \, x_{i \lambda}^4 
x_{j \lambda}^4 \, >>.
\end{equation} 
This follows from summing over the index $\mu$. 

Further using the symmetry in the form of $S$ gives
\begin{equation}
a_{1}+(N-1)a_{2}+3(N-1)(N-2)a_{3}=0
\end{equation}
Thus putting together this information leads to 
\begin{equation}
\frac{S(f=l/N)}{N(N-1)}
=f(1-f)\left[ \frac{a_{1}N^2}{(N-1)}-N^2(N-1)a_{4} \right] +f^2 (1-f)^2 a_{4}N^4.
\end{equation}
The symmetry about $f=1/2$ is now obvious.  $a_{1}$ can be evaluated
exactly using the methods in \cite{Pandey} and gives
\begin{equation}
a_{1}=9/(N(N+2)(N+4)(N+6)) \sim 9/N^4.
\end{equation}
The correlation $a_{4}$ is unfortunately beyond the author, but it is
sufficient to observe that
\begin{equation}
a_{4} \sim 3 <<\, x_{i \lambda} x_{j \lambda} x_{i \mu} x_{j \mu} \, >>^2 = 3/N^6
\end{equation}
The corrections to this are of order $N^{-7}$. The factor 3 is neccessary as the number of 
different ways of choosing 2 indices from 4 is 6 and choosing any two results in the automatic
choice of the other pair. This new two-vector correlation is easy to evaluate (\cite{Pandey}) and we 
get the above result. 

Collecting all the correlations we finally arrive at 
\begin{equation}
S(f=l/N)\, =\, 3 f^2 (1-f)^2 + O(N^{-1})
\end{equation}
Thus we have proved that $S$ is only a function of $f$ in the large $N$ limit. This 
functional form should match those for time-reversal symmetric systems with no special 
symmetries. This is illustrated in Fig. 6, and the fit is satisfactory. 
The standard deviation of the fluctuations is for such systems thus given simply
by 
\begin{equation}
\sigma = \sqrt{3}/N + O(N^{-2}),
\end{equation}
and implies the independence of the width of the fluctuations from the size of the 
partitions. 
In contrast the numerical results show a strong $f$ dependence for systems with
special symmetries such as parity \cite{LakPRE}. Further work is underway for 
elucidating features of the time-reversal breaking case as well.

\section{Summary}
We have studied correlations between operators in  quantum chaotic
systems. The relaxation towards an oscillatory state of decorrelation
or ``equilibrium'' has been analysed numerically and also from the 
point of view of RMT. Certain universal features have been found 
and in the case of time-reversal symmetric systems these have been
derived using RMT. 

There are many natural extensions to this work, the most important 
being consideration of general density operators rather than projectors.	Further it has been found in \cite{LakPRE} that the relaxation
	process is Gaussian implying that the mean and standard
	deviation characterise the process completely. 
The question of energy has not arisen in this work because of 
	our models. Therefore the generalization of this work
	to Hamiltonian flows with infinite Hilbert spaces is of 
	considerable importance. 
Connections have to be made between these approaches in quantum 
chaos and the existing methods of quantum statistical mechanics.

	The effect of scarring of wavefunctions by classical 
periodic orbits tends to localize many quantum states. It would be
of interest to see if the measures defined herein are sensitive to
this phenomenon. We have already noted deviations from RMT in the case 
of the quantized bakers map which is known to have a large number 
of scarred eigenstates.

\newpage

\begin{figure}[h]
\hspace*{.5in}\psfig{figure=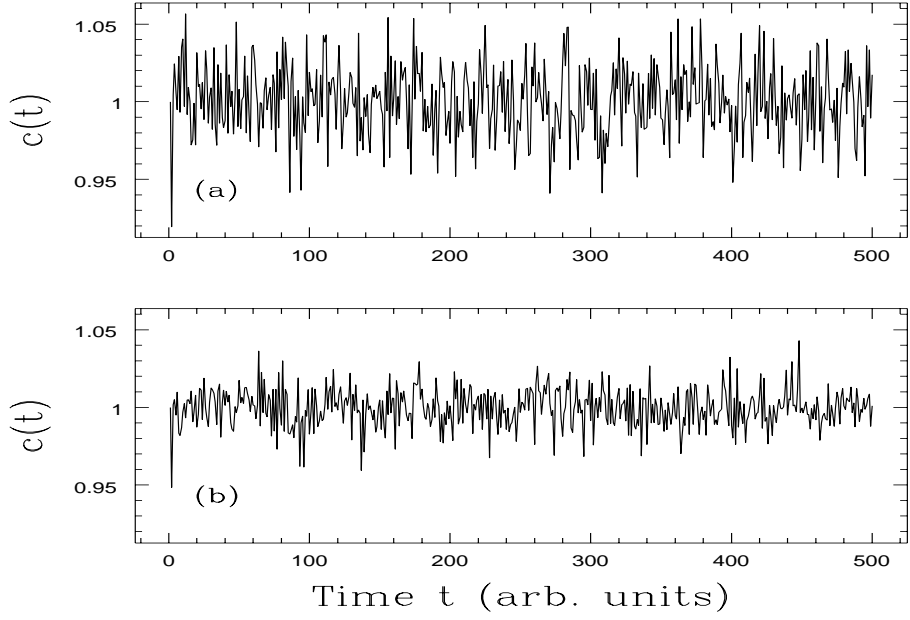,height=3.5in,width=5in}
\vspace{0.1in}
\caption{The relaxation fluctuations as a function of time for the
symmetric standard map with $K=20$ and (a) $N=100$ (b) $N=200$.}
\label{fig:1}
\end{figure}

\begin{figure}[h]
\hspace*{.5in}\psfig{figure=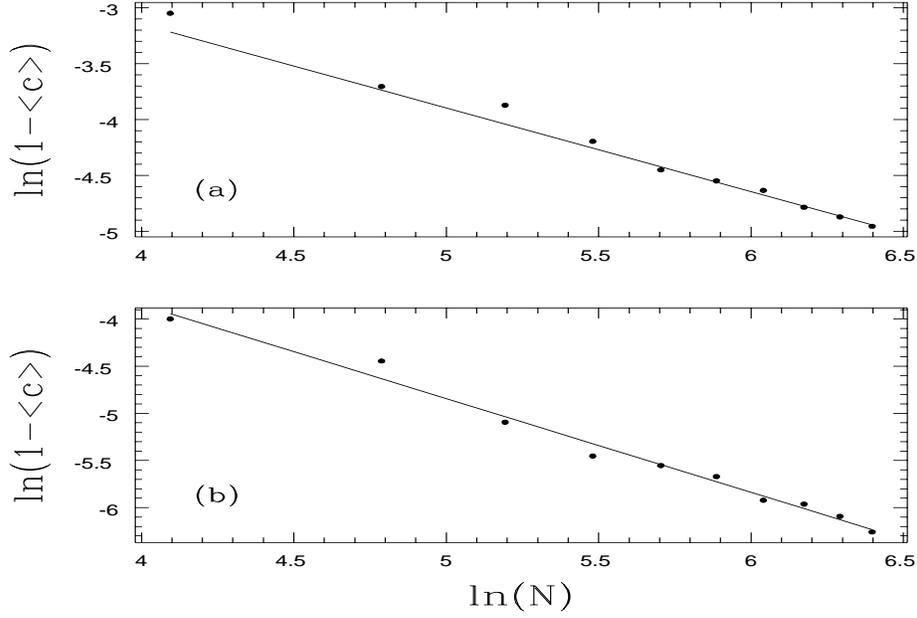,height=3.5in,width=5in}
\vspace{0.4cm}
\caption{ The average of the fluctuations for (a) the quantum bakers
map quantizing the (2/3,1/3) Bernoulli scheme and (b) the standard map with $K=20$. Shown 
is the logarithm of the
deviation from unity  as a function of $\log(N)$.}
\label{fig:2}
\end{figure}

\begin{figure}[h]
\hspace*{.5in}\psfig{figure=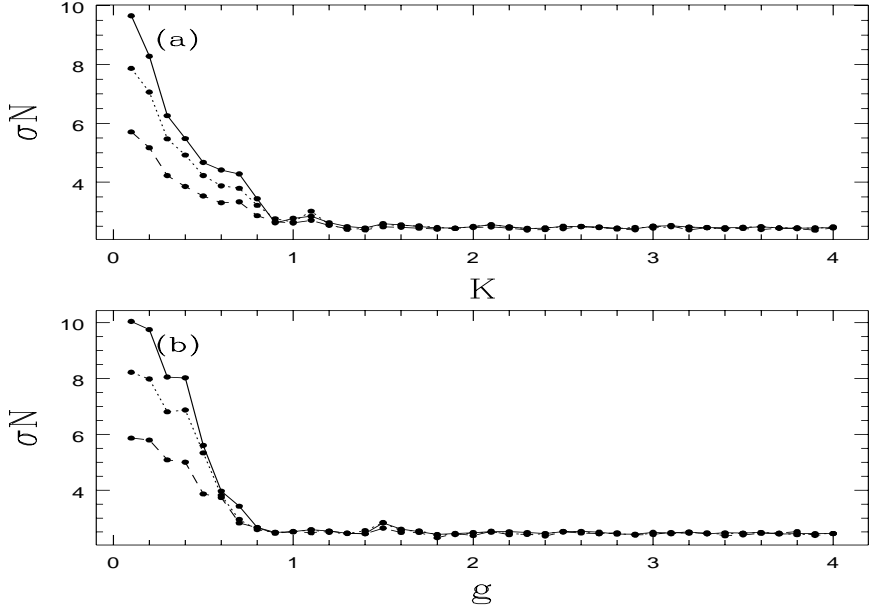,height=3.5in,width=5in}
\vspace{0.4cm}
\caption{The scaled standard deviation $\sigma N$ as a function of the
kick strength for (a) the symmetric standard map, (b) the symmetric
Harper map. The solid line corresponds
to $N=300$, the dotted line to N=200 and the dashed line to N=100.}
\label{fig:3}
\end{figure}

\begin{figure}[h]
\hspace*{.5in}\psfig{figure=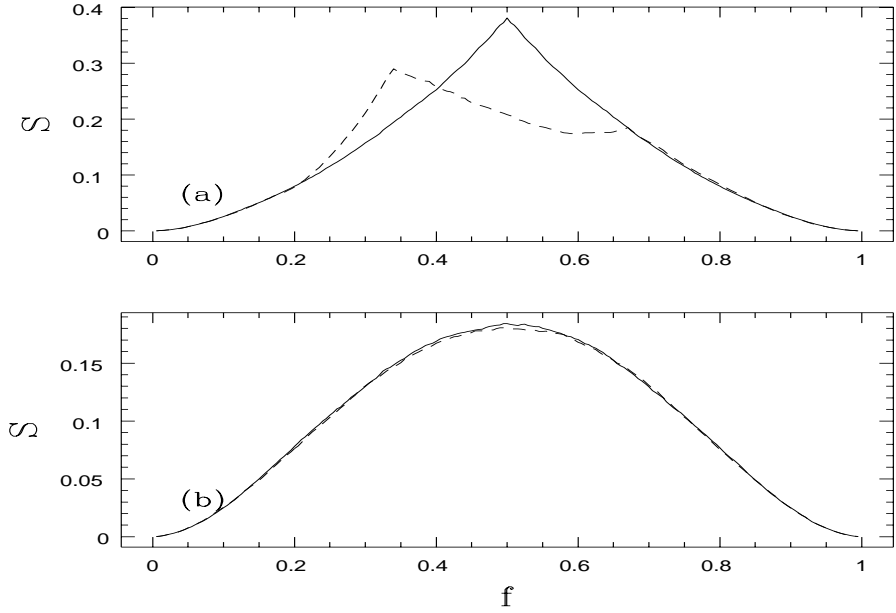,height=3.5in,width=5in}
\vspace{0.4cm}
\caption{The eigenfunction sum $S$ as a function of $f$. (a) The symmetric 
standard map with $N=200$, the solid line is when $n_{0}=0$ 
and the dashed one for
$n_{0}=66$. (b) Case of the asymmetric standard map with $(a=0.35,b=0.0)$,
the solid line is when $n_{0}=0$ and the dashed one for $n_{0}=66$. In
all cases $K=20$.}
\label{fig:4}
\end{figure}

\begin{figure}[h]
\hspace*{.5in}\psfig{figure=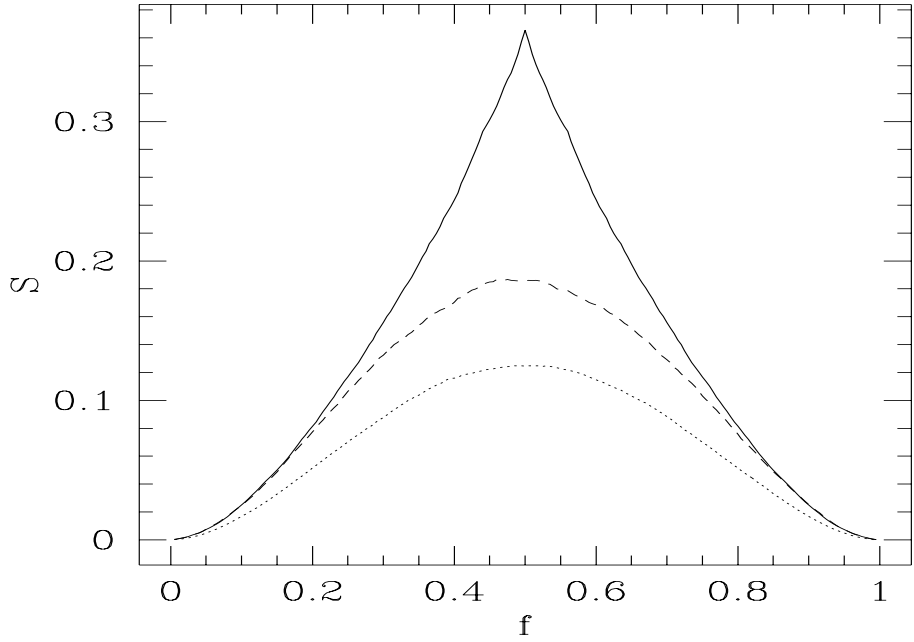,height=3.5in,width=5in}
\vspace{0.4cm}
\caption{The eigenfunction sum $S$ for the Harper map with $g=8$. The
symmetric Harper map with $(a=1/2,b=1/2)$ is the solid line while
the case $(a=0.35,b=1/2)$ is the time-reversal symmetric case with no
special phase space symmetries and the dotted line $(a=0.35,b=0.35)$
breaks also time-reversal symmetry. In all cases $N=200$.}
\label{fig:5}
\end{figure}

\begin{figure}[h]
\hspace*{.5in}\psfig{figure=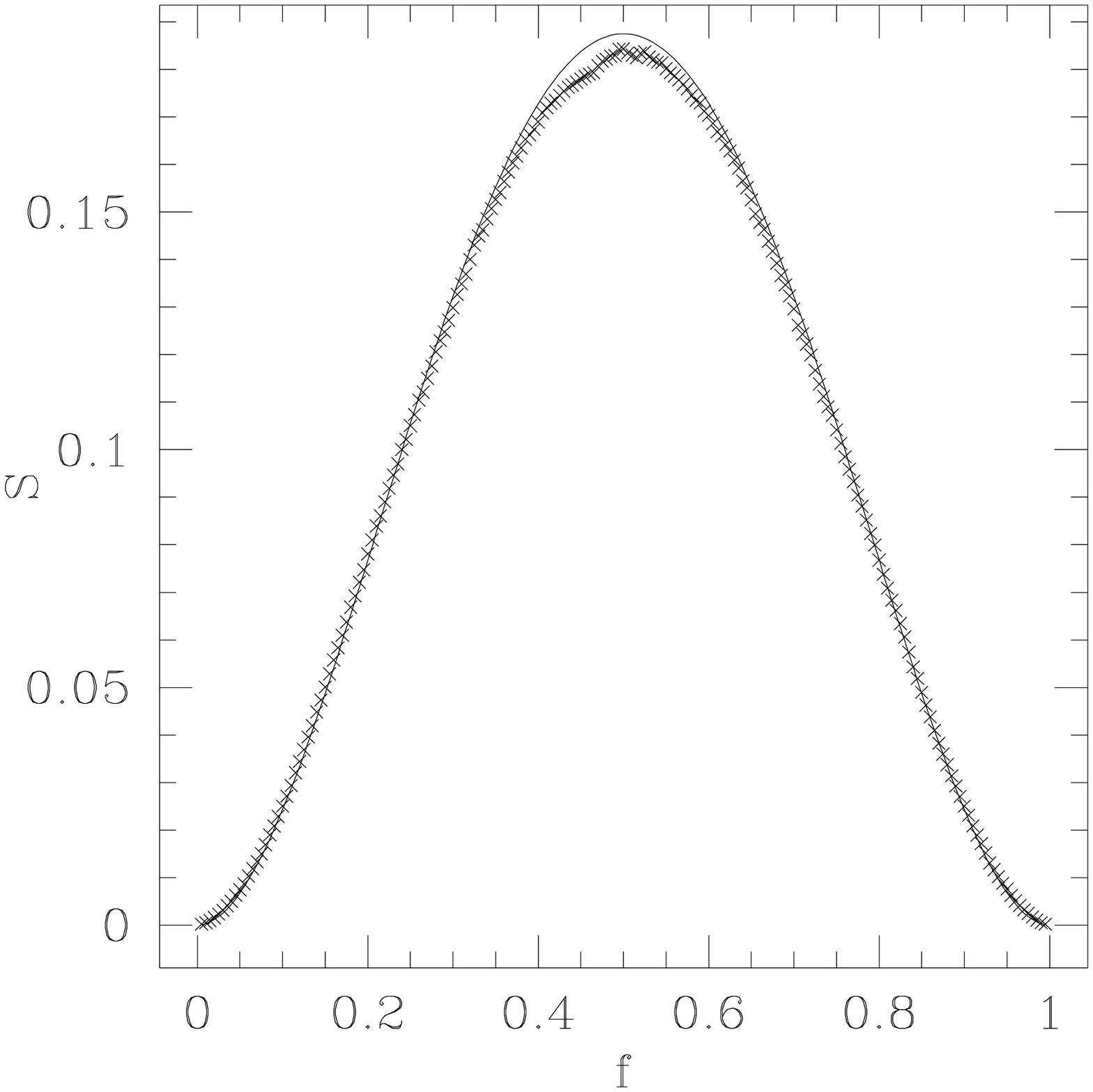,height=3.5in,width=5in}
\vspace{0.4cm}
\caption{Comparison of the prediction from RMT, the solid line, with the 
asymmetric standard map (N=200,K=20,a=0.35,b=0), the crosses.}
\label{fig:6}
\end{figure}

\begin{references}
\bibitem[*]{byline} e-mail: arul@prl.ernet.in

\bibitem{ArnAvez} V. I. Arnold and A. Avez, {\it Ergodic Problems of
Classical Mechanics} WA Benjamin, New York (1968).
\bibitem{Weigert} S.Weigert in {\it Adriatico Research Conference and
Miniworkshop, Quantum Chaos}, edited by H. A. Cerdeira, R. Ramaswamy,
M. C. Gutzwiller and G. Casati (World Scientific, 1990).
\bibitem{Izrailev}  F. M. Izrailev, Phys. Rep. {\bf 196}, 299 (1990).
\bibitem{ElsKap} Y. Elskens and R. Kapral, J. Stat. Phys. {\bf 38}, 1027 (1985).
\bibitem{GaspJsp92} P. Gaspard, J. Stat. Phys. {\bf 68}, 673 (1992).
\bibitem{UBT} O. Bohigas, S. Tomsovic, and D. Ullmo, Phys. Rep. {\bf 233},
45 (1993).
\bibitem{LBJsp} A. Lakshminarayan, and N. L. Balazs, J. Stat. Phys. {\bf 77},
311 (1994).
\bibitem{LakPRE} A. Lakshminarayan, Phys. Rev. E. {\bf 56}, 2540 (1997) (chao-dyn/9704001).
\bibitem{SurvProb} F. Leyvraz {\it et. al.}, Phys. Rev. Lett. {\bf 67}, 2921 (1991); 
Y. Alhassid and N. Whelan, Phys. Rev. Lett. {\bf 70}, 572 (1993); A. Tameshtit and 
J. E. Sipe, Phys. Rev. A {\bf 45}, 8280 (1992); J. Wilkie and P. Brumer, Phys. Rev. Lett. {\bf 67},
1185 (1991).  
\bibitem{Leb} P. Leboeuf {\it et. al.}, Phys. Rev. Lett. {\bf 65}, 3076 (1990).
\bibitem{BVbak} N. L. Balazs, and A. Voros, Ann. Phys. (N. Y.) {\bf 190},
1 (1989).
\bibitem{Sarabak} M. Saraceno, Ann. Phys. (N.Y.) {\bf 199}, 37 (1990).
\bibitem{Pandey} T. A. Brody, {\it et. al. } Rev. Mod. Phys., {53} 385 (1981).
\bibitem{SteveHeller} P. W. O'Connor, S. Tomsovic, and E. J. Heller, Physica 
{\bf 55D}, 340 (1992).   
\bibitem{ShiChang} S.J. Chang, and K. J. Shi, Phys. Rev. Lett. {\bf 55},
269 (1985).

\end{references}
\end{document}